\documentclass[11pt,a4paper]{article}
\usepackage{authblk}
\usepackage[utf8]{inputenc}
\usepackage[round]{natbib}
\usepackage[pdftex]{graphicx}
\usepackage{amsmath,amsthm,amsfonts}
\usepackage{lscape}   
\usepackage{rotating}
\usepackage{rotfloat}
\usepackage{multirow}
\usepackage{booktabs}
\usepackage{ctable}
\usepackage{threeparttable}
\usepackage{booktabs}
\usepackage{framed}
\usepackage{color}
\usepackage[colorlinks=true,linkcolor=blue,allcolors=blue]{hyperref}

\usepackage{soul}
\usepackage[ruled,vlined]{algorithm2e}

\title{Wavelet-based estimation in aggregated functional data with positive and correlated errors}

\author{Alex Rodrigo dos Santos Sousa \thanks{asousa@unicamp.br}}
\author{João Victor Siqueira Rodrigues}
\author{Vitor Ribas Perrone}
\author{Raul Gomes Rocha}

\affil{Universidade Estadual de Campinas (UNICAMP)\\ Departament of Statistics, Brazil}

\date{} 

\begin{document}

\maketitle

\vspace{-1.0cm}
\begin{abstract}
  We consider the statistical problem of estimating constituent curves from observations of their aggregated curves, referred to as aggregated functional data, in models with strictly positive random errors under Gamma distribution and correlated errors under AR(1) and ARFIMA processes. This problem arises in several areas of knowledge, such as chemometrics, for example, when absorbance curves of the constituents of a given substance must be estimated from its aggregated absorbance curve according to the Beer–Lambert law.

In this sense, we propose Bayesian wavelet-based methods to estimate the component functions under a functional data analysis approach. This procedure has the advantage of estimating curves with important local features, such as discontinuities, peaks, and oscillations, due to the expansion properties of functions in wavelet bases. We further evaluate the performance of the proposed method through computational simulations, as well as applications to real data. \\
  
  \end{abstract}



\maketitle

\section{Introduction}
Across several scientific domains, one encounters aggregated functional data generated by unknown component curves. In spectroscopy, for instance, there is interest in estimating the individual mean absorbance curves of the constituents of a given substance from samples of aggregated absorbance curves of the substance itself. In this setting, by the Beer–Lambert law (\cite{brereton2003}), the overall absorbance is expressed as a convex linear combination of the absorbances of its constituents, with the combination coefficients corresponding to their concentrations in the samples. This problem is commonly referred to as calibration in chemometrics. Another example arises in modeling the composition of electricity consumption in a given region, where the average load curve over a specified time interval is formed by aggregating the consumption curves of individual consumers. For further details on these examples, see \cite{dias-2009} and \cite{dias-2013}.

A variety of statistical methodologies have been proposed to estimate component curves from aggregated observations. Many of these approaches treat the data as multivariate because measurements are collected at discrete time points. Thus, each sampled point on the aggregated curve is considered as a variable with an associated correlation structure. Principal Component Regression (PCR) by \cite{cowe1985} and Partial Least Squares (PLS) by \cite{sjostrom1983} are usual multivariate methodologies for handling aggregated data. Bayesian approaches and wavelet-based methods were also proposed by \cite{brown1998a,brown1998b,brown2001}. Although successful in many applications, such methods do not explicitly account for the intrinsic functional structure of the data, despite the fact that discretization arises merely from instrumental constraints, see \cite{ramsay1997} for an overview of functional data analysis. In this direction, successful functional approaches were first developed by \cite{dias-2009,dias-2013}, employing spline or B-spline expansions. However, while spline expansions perform well for smooth curves, their performance deteriorates when the underlying component functions exhibit localized features such as discontinuities, sharp peaks, or oscillatory behavior.
In this sense, \cite{sousa-2023,sousa-2024} proposed a wavelet-based approach in functional data analysis to estimate component functions in aggregated curves. 

However, both the multivariate and functional models typically assume additive Gaussian errors, even though non-Gaussian noise is frequently observed in practice. For example, \cite{huppert2016} and \cite{sung2019} discuss the occurrence of non-Gaussian noise in spectroscopic measurements. Motivated by these limitations, we propose the use of wavelet-based methods to estimate individual mean component curves from aggregated observations under models with two scenarios: (i) strictly positive additive noise with Gamma distribution and (ii) correlated noise under an autoregressive or ARFIMA process. In addition, we propose to employ Bayesian wavelet shrinkage rules based on a mixture prior consisting of a point mass at zero and either a logistic distribution, as introduced by \cite{sousa-2022}, to estimate wavelet coefficients from empirical coefficients obtained via the discrete wavelet transform (DWT) applied to the aggregated data. 

Additive strictly positive errors in the original model introduce substantial inferential challenges. First, the independence property is generally lost after the wavelet transform, that is, the transformed errors in the wavelet domain are typically correlated. Consequently, wavelet coefficients cannot be estimated independently, as is commonly done under Gaussian noise assumptions, but must instead be inferred jointly from their posterior distribution. This requirement typically necessitates computational strategies such as Markov Chain Monte Carlo (MCMC) methods for sampling from the joint posterior distribution. Furthermore, although the original errors are strictly positive, their wavelet-domain representations correspond to linear combinations and are not necessarily positive themselves, see \cite{sousa-2023b}. While several statistical models with positive multiplicative noise have been addressed via logarithmic transformations, models with positive additive noise are comparatively rare in the literature, despite their practical relevance. See \cite{radnosrati2020} for a development of estimation theory in models with positive noise.

On the other hand, correlated noises offer a different challenge. Although the decorrelation property of the DWT, the variabilities of the wavelet coefficients are different among resolution levels. Thus, the estimation process requires a level-dependent approach that takes this difference into account, see \cite{johnstone-silerman-1997}. 

This paper is organized as follows: Section 2 presents the statistical models involving aggregated curves considered in this work. The wavelet-based estimation procedures are described in Section 3. Simulation studies to evaluate the performance of the proposed methods are analyzed and discussed in Section 4. Real data applications are discussed in Section 5. Finally, Section 6 provides final considerations.

\section{Statistical models}
We consider a function $A(t)$ composed of a convex linear combination of $L \geq 2$ functions according to the model.
\begin{equation} \label{func_model}
    A(t) = \sum_{l=1}^{L} y_l \alpha_l(t) + \epsilon(t),
\end{equation}
where $t \in \mathbb{R}$, $\alpha_l(t) \in \mathbb{L}_2(\mathbb{R}) = \{f(t):\int_\mathbb{R} f^2(t)dt < \infty\}$ are unknown squared integrable functions, $y_l \in (0,1)$ and $\sum_{l=1}^{L} y_l = 1$ are known weights, $l = 1,\ldots,L$ and $\epsilon(t)$ is a random error process. We will call $A(t)$ the aggregated function and $\alpha_l(t)$ the component functions of the aggregation. 

In practice, we observe a sample of $N$ aggregated curves at $M = 2^J (J \in \mathbb{N})$ equally spaced locations according to the discrete version of the model \eqref{func_model},
\begin{equation} \label{discr_model}
    A_n(t_m) = \sum_{l=1}^{L} y_{nl} \alpha_l(t_m) + \epsilon_n(t_m),
\end{equation}
$n = 1,\ldots,N$ and $m = 1,\ldots,M$. The goal is to estimate the functions of the $L$ components $\alpha_l(t)$ using observations $(t_m,A_n(t_m))$ as data. This step is known as the calibration of the model \eqref{func_model}. 

Functional approaches to the calibration problem when $\epsilon(t)$ is a Gaussian process were proposed first by \cite{dias-2009} and \cite{dias-2013} using splines expansions of the component functions and recently by \cite{sousa-2023} and \cite{sousa-2024} using wavelets representations of them. In this work, we consider the wavelet-based estimation of the component functions when the random errors are not Gaussian. Specifically, two scenarios are considered on the distribution of the random errors:
\begin{enumerate}
    \item $\epsilon_n(t_m)$ are positive, independent and identically distributed (iid) and have Gamma distribution with parameters $a > 0$ and $b > 0$, i.e, $\epsilon_n(t_m) \sim \mathrm{Gamma}(a,b)$ and has probability density function given by (we dropped the indices for simplicity),
    \begin{equation} \nonumber
        h(\epsilon) = \frac{b^a}{\Gamma(a)}\epsilon^{a-1}e^{-b\epsilon}\mathbb{I}(\epsilon>0),
    \end{equation}
    where $\Gamma(\cdot)$ and $\mathbb{I}(\cdot)$ are the Gamma and the indicator functions respectively.

    \item $\epsilon_n(t_m)$ are correlated and follow: (i) an autoregressive (AR) process of order 1, i.e, $\epsilon_n(t_m) \sim \mathrm{AR}(1)$,
    \begin{equation} \nonumber
        \epsilon_i = \phi \epsilon_{i-1} + \eta_i,
    \end{equation}
    where $\phi \in (-1,1)$ and $\eta_i$ are iid with Gaussian distribution, $\eta_i \sim \mathrm{N}(0,\sigma^2_{\eta})$, $\sigma_\eta > 0$ and (ii) an autoregressive fractionally integrated moving-average process (ARFIMA) of order $(0,d,0)$, i.e, $\epsilon_n(t_m) \sim \mathrm{ARFIMA}(0,d,0)$, $0 < d < 0.5$,
\begin{equation} \nonumber
    (1-B)^d \epsilon = \eta_i,
\end{equation}
where $B$ is the backshift operator, $(1-B)^d=\sum_{q=0}^\infty b_q(d)B^q$ with $b_q(d) = \Gamma(q-d)/(\Gamma(q+1)\Gamma(-d))$ for $q=0,1,\ldots$   
\end{enumerate}

The scenarios have an important impact on the estimation procedure under the wavelet approach. If the random errors in the time domain are iid zero-mean Gaussian, then the random errors in the wavelet domain remain iid zero-mean Gaussian with the same scale parameter. However, if the errors are not Gaussian, the distribution of their counterparts in the wavelet domain is not preserved, see \cite{neumann-1995}.

\section{Estimation procedures}
The estimation process starts by taking the data from the time domain to the wavelet domain. The discretized model \eqref{discr_model} can be written in matrix notation as
\begin{equation} \label{matrix_model}
    \boldsymbol{A} = \boldsymbol{\alpha} \boldsymbol{y} + \boldsymbol{\epsilon}, 
\end{equation}
where $\boldsymbol{A} = (A_n(t_m))_{M \times N}$ is the $M \times N$ matrix with the aggregated observations in which each column is a sample of $M$ points, $\boldsymbol{\alpha} = (\alpha_l(t_m))_{M \times L}$ is the $M \times L$ matrix with the unknown component functions, $\boldsymbol{y} = (y_{nl})_{L \times N}$ is the $L \times N$ matrix with the weights of the component functions and $\boldsymbol{\epsilon} = (\epsilon_n(t_m))_{M \times N}$ is the $M \times N$ matrix with the random errors. We apply a Discrete Wavelet Transform (DWT) on the data, which can be represented by the multiplication of an $M \times M$ orthogonal transformation matrix $\boldsymbol{W}$ on both sides of \eqref{matrix_model}, obtaining the model in the wavelet domain,
\begin{equation} \label{wav_model}
    \boldsymbol{D} = \boldsymbol{\Theta} \boldsymbol{y} + \boldsymbol{\varepsilon},
\end{equation}
where $\boldsymbol{D} = \boldsymbol{W}\boldsymbol{A}$ is the $M \times N$ matrix with the empirical (observed) aggregated wavelet coefficients, $\boldsymbol{\Theta} = \boldsymbol{W}\boldsymbol{\alpha}$ is the $M \times L$ matrix with the unknown wavelet coefficients of the component functions and $\boldsymbol{\varepsilon} = \boldsymbol{W}\boldsymbol{\epsilon}$ is the $M \times N$ matrix with the random errors in the wavelet domain. To remove the noise in \eqref{wav_model}, we apply a wavelet shrinkage rule $\boldsymbol{\delta}(\cdot)$ in the matrix with the empirical coefficients $\boldsymbol{D}$ and estimate $\boldsymbol{\Theta}$ by 
\begin{equation} \label{est_theta}
    \hat{\boldsymbol{\Theta}} = \boldsymbol{\delta}(\boldsymbol{D)}\boldsymbol{y'}(\boldsymbol{yy'})^{-1}.
\end{equation}
Finally, the matrix $\boldsymbol{\alpha}$ is estimated by applying the Inverse Discrete Wavelet Transform (IDWT), represented by the transpose matrix of $\boldsymbol{W}$,
\begin{equation} 
  \hat{\boldsymbol{\alpha}} = \boldsymbol{W'} \hat{\boldsymbol{\Theta}}. \nonumber
\end{equation}
See \cite{sousa-2024} for more details.

The distribution of random errors in \eqref{discr_model} affects the wavelet shrinkage rule to be applied to the empirical wavelet coefficients in \eqref{est_theta}. The next subsections provide details of the wavelet shrinkage procedure under the two scenarios of errors considered.

\subsection{Application of wavelet shrinkage under positive (Gamma) random error}

Under the scenario of iid positive random errors with Gamma distribution in the model \eqref{discr_model}, we have that the probability density function of $\boldsymbol{\varepsilon} = W\epsilon$ is given by
\begin{equation}
    f(\boldsymbol{\varepsilon}) = \left(\frac{b^a}{\Gamma(a)}\right)^n \exp\left\{-b \sum_{i=1}^n\sum_{k=1}^n w_{ki} \varepsilon_k\right\} \left( \prod_{i=1}^n\sum_{k=1}^n w_{ki} \varepsilon_k \right)^{a - 1} \prod_{i=1}^n \mathbb{I}_{(0,\infty)}\left(\sum_{k=1}^nw_{ki}\varepsilon_k\right). \nonumber
\end{equation}
   Then, doing $\boldsymbol{\varepsilon} = \boldsymbol{d} - \boldsymbol{\theta}$, the likelihood function is   
\begin{align}\label{lik_pos}
    \mathcal{L}(\boldsymbol{\theta} \mid \boldsymbol{d}) = &\left(\frac{b^a}{\Gamma(a)}\right)^n \exp\left\{-b \sum_{i=1}^n\sum_{k=1}^n w_{ki} (d_k -\theta_k)\right\} \left( \prod_{i=1}^n\sum_{k=1}^n w_{ki} (d_k -\theta_k) \right)^{a - 1} \times \nonumber \\&\times\prod_{i=1}^n\mathbb{I}_{(0,\infty)}\left(\sum_{k=1}^n w_{ki} (d_k -\theta_k)\right).
\end{align}
    
We adopt the mixture of a point mass function at zero and the centered-around-zero logistic distribution with scale parameter $\tau >0$ as prior distribution to a single wavelet coefficient, as proposed by \cite{sousa-2022}. Thus, the joint prior distribution of $\boldsymbol{\theta}$ is
\begin{equation}\label{prior_pos}
    \pi(\boldsymbol{\theta}) = \prod_{i=1}^n[p \delta_0(\theta_i) + (1 - p) g(\theta_i; \tau),
\end{equation}
where $p \in (0,1)$, $\delta_0(\cdot)$ is the point mass function at zero and $g(\cdot; \tau)$ is the logistic probability density function centered at zero. Considering \eqref{lik_pos} and \eqref{prior_pos}, the associated posterior distribution is 
\begin{align} \label{post}
    \pi(\boldsymbol{\theta} \mid \boldsymbol{d}) & \propto  \pi(\boldsymbol{\theta}) \mathcal{L}(\boldsymbol{\theta} \mid \boldsymbol{d}) \nonumber \\ 
    &\propto \left(\prod_{i=1}^n \left[ p \delta_0(\theta_i) + (1 - p) \frac{\exp\left\{-\frac{\theta_i}{\tau}\right\}}{\tau \left( 1 + \exp\left\{-\frac{\theta_i}{\tau}\right\} \right)^2} \right]\right) \exp\left\{-b \sum_{i=1}^n\sum_{k=1}^n w_{ki} (d_k -\theta_k)\right\} \times  \nonumber \\
    & \times \left( \prod_{i=1}^n\sum_{k=1}^n w_{ki} (d_k -\theta_k) \right)^{a - 1}\prod_{i=1}^n\mathbb{I}_{(0,\infty)}\left(\sum_{k=1}^n w_{ki} (d_k -\theta_k)\right).
\end{align}
Note that, although the error $\boldsymbol{\epsilon}$ in the time domain is independent, after applying the DWT the error in the wavelet domain $\boldsymbol{ \varepsilon}$ is no longer independent, a fact due to the loss of normality. Therefore, the shrinkage rule must be applied to the vector of coefficients, and it is no longer possible to apply it coefficient by coefficient.

In Bayesian settings, the shrinkage rule is computed through the posterior expectation, that is, $\delta(\boldsymbol{d}) = \mathbb{E}(\boldsymbol{\theta} \mid \boldsymbol{d})$. However, in this case this is not analytically tractable, requiring the use of Markov Chain Monte Carlo (MCMC) methods. The idea is to generate $K$ samples from the posterior distribution \eqref{post} and approximate $\hat{\theta}_i$ by the average of the generated samples, that is,
\[
\hat\theta_i = \delta_i(\boldsymbol{d}) \approx\frac{1}{K} \sum_{k=1}^K \theta_{ki}.
\]
It is worth noting that, when using this method, it is common to employ a \textit{burn-in} period, in which the initial part of the chain is discarded to reduce dependence on the starting point, and thinning, which consists of retaining only every $j$-th sample generated by the chain in order to reduce the autocorrelation among successive iterations. See \cite{casella-2004} for more details about MCMC methods.

To approximate this expectation, we employ the Algorithm \ref{alg:RAM}, the Robust Adaptive Metropolis (RAM) algorithm proposed by \cite{vihola-2012}. This method relies on a lower triangular matrix $\boldsymbol{S}_k$ with positive diagonal elements, which is updated at each iteration. Moreover, $\{\eta_l\}_{l \geq 1} \subset (0,1]$ denotes a sequence that decays to zero, and $\gamma \in (0,1)$ represents the target average acceptance rate. 

To elicite the hyperparameters $\tau$ and $p$ in \eqref{prior_pos}, we follow the proposals of \cite{sousa-2022} and \cite{angelini-vidakovic-2004} respectively, i.e, $\tau$ should be chosen such that $0 <\tau \leq 10$ and   
\begin{equation}\label{alpha}
p = p(j) = 1 - \frac{1}{(j-J_{0}+1)^h},
\end{equation}
where $J_ 0 \leq j \leq J-1$, $J_0$ is the primary resolution level, $J$ is the number of resolution levels, $J=\log_{2}(n)$ and $h > 0$. They also suggested that in the absence of additional information, $h = 2$ can be adopted.

\begin{algorithm}[htb!]
\caption{Robust Adaptive Metropolis (RAM).}\label{alg:RAM}
\DontPrintSemicolon
\KwIn{$\boldsymbol{\theta}_{1}$, $\boldsymbol{S}_1$, $\{\eta_l\}_{l \geq 1} \subset (0,1]$, $\gamma \in (0,1)$, and $K$ (number of iterations).}
\KwOut{Samples $\boldsymbol{\theta}_1, \dots, \boldsymbol{\theta}_K$ from $\pi (\boldsymbol{\theta} \mid \boldsymbol{d})$.}
\BlankLine
$k \gets 2$\;
Generate $\boldsymbol{\theta}^*_k = \boldsymbol{\theta}_{k-1} + \boldsymbol{S}_{k-1} \boldsymbol{U}_k$, where $\boldsymbol{U}_k \sim N_n(\boldsymbol{0}, \boldsymbol{I})$\;
Set $\boldsymbol{\theta}_k = \boldsymbol{\theta}_k^*$ with probability $\gamma_k$, otherwise set $\boldsymbol{\theta}_k = \boldsymbol{\theta}_{k-1}$, where
$\gamma_k = \min\left( 1, \dfrac{\pi(\boldsymbol{\theta}_k^*)}{\pi(\boldsymbol{\theta}_{k-1})} \right)$\;
Compute the lower triangular matrix $\boldsymbol{S}_k$ with positive diagonal elements satisfying
\[
\boldsymbol{S}_k \boldsymbol{S}_k' = \boldsymbol{S}_{k-1} \left(\boldsymbol{I} + \eta_k (\gamma_k - \gamma) \frac{\boldsymbol{U}_k \boldsymbol{U}_k'}{\|\boldsymbol{U}_k\|^2} \right) \boldsymbol{S}_{k-1}'
\]\;
$k \gets k + 1$ and, if $k \leq K$, return to Step 2\;
\end{algorithm}

\subsection{Application of wavelet shrinkage under correlated random error}
In the second scenario, the errors in \eqref{discr_model} are assumed to be correlated, following $\mathrm{AR}(1)$ and $\mathrm{ARFIMA}(0,d,0)$ processes. Due to the decorrelation property of the DWT, the shrinkage rule can be applied individually to each empirical coefficient. However, applying the rule involves estimating the standard deviation of the empirical coefficients at each resolution level, i.e, the shrinkage rule must be level-dependent, see \cite{johnstone-silerman-1997}. In this sense, we apply the Bayesian approach of \cite{sousa-2025} to the context of aggregated functional data. The shrinkage rule to be applied on a single empirical wavelet coefficient $d$ at the resolution level $j$ is
\begin{equation}\label{cor_rule}
    \delta_j(d) = \displaystyle \frac{(1 - p) \int_{\mathbb{R}} (\hat{\sigma}_j u + d) g(\hat{\sigma}_j u + d; \tau) \phi(u) du}{\frac{p}{\hat{\sigma}_j} \phi\left( \frac{d}{\hat{\sigma}_j} \right) + (1 - p) \int_{\mathbb{R}} g(\hat{\sigma}_j u + d; \tau) \phi(u) du}, \nonumber
\end{equation}
where
\begin{equation}\label{cor_sd}
    \hat{\sigma}_j = \hat{\sigma}(j) = \displaystyle \frac{\operatorname{median}\{\, |d_{j,k}| : k = 0, \ldots, 2^{j} \,\}}{0.6745}. \nonumber
\end{equation}

\section{Simulation studies}
\begin{figure}[htb!]
    \centering
    \includegraphics[width=0.85\linewidth]{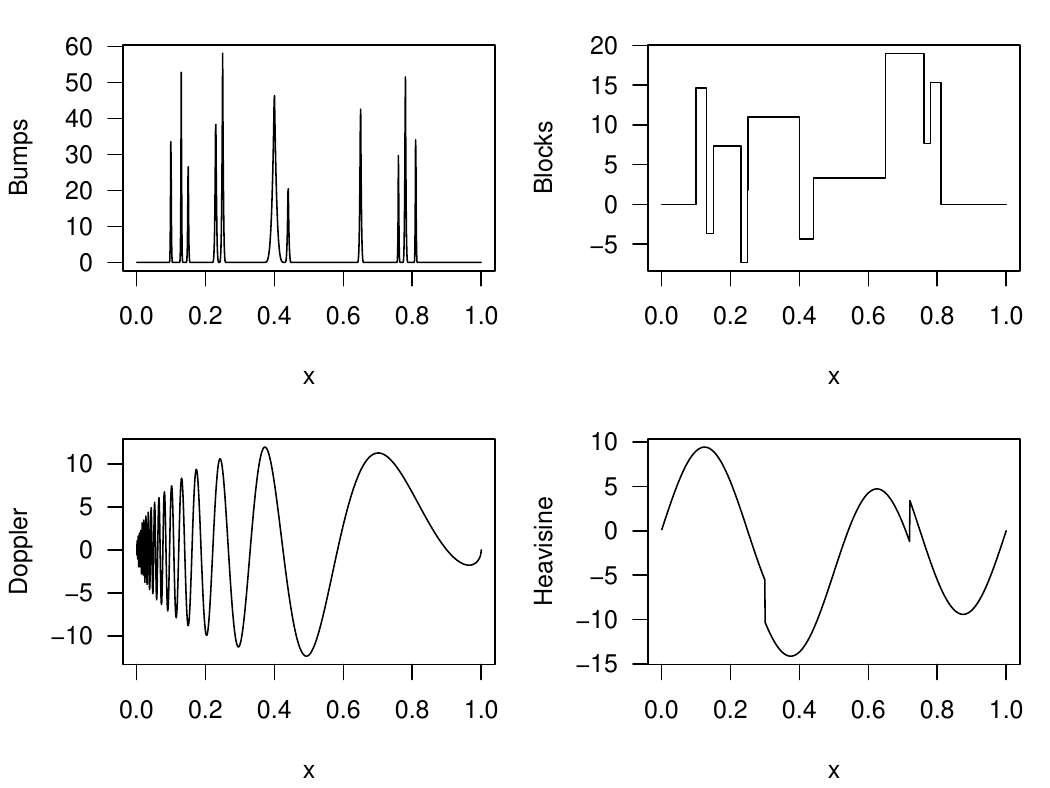}
    \caption{Donoho and Johnstone test functions.}
    \label{fig: dj_functions}
\end{figure}

We conducted simulation studies to evaluate the performance of the proposed methods. 
To do so, we adopted the so-called Donoho and Johnstone (DJ) test functions as underlying component functions in \eqref{func_model}. These functions called Bumps, Blocks, Doppler and Heavisine have interesting local features to be estimated such as peaks, discontinuities and oscillations, see \cite{DJ-1994} for more details. Figure \ref{fig: dj_functions} shows the plots of the DJ-test functions.

In both positive and correlated errors scenarios, we generated the data according to the model \eqref{discr_model} for specific values of sample size $N$, number of points $M$ and number of component functions $L$. Further, the variance of the error was also obtained according to specific values of signal-to-noise ratio (SNR). As performance measure, for each replication $r$ and component function $\alpha(\cdot)$, we calculated the mean squared error (MSE),
\begin{equation} \nonumber
 \mathrm{MSE}_r = \frac{1}{M} \sum_{i=1}^{M} \left( \hat{\alpha}^{(r)}(t_i) - \alpha(t_i) \right)^2, 
 \end{equation}
\noindent where \(\hat{\alpha}^{(r)}(t_i)\) denotes the estimate of $\alpha(t_i)$, for \(i = 1, \dots, M\). Then, we finally obtained the averaged mean squared error (AMSE)
\begin{equation} \nonumber
\mathrm{AMSE} = \frac{1}{R} \sum_{r=1}^{R} \mathrm{MSE}_r,  
\end{equation}
where $R$ is the number of replications.

In the next two subsections we present more details and the results for each error scenario.

\subsection{Results for positive (Gamma) error}
In the model \eqref{discr_model} under iid gamma errors, we adopted $N = 50$ curves, $M = 32$ and $64$ equally spaced points, $L = 2$ (Bumps and Doppler as underlying component functions) and $4$ (all the DJ test functions as underlying component functions) and $\mathrm{SNR} = 3$ and $7$. Further, we chose a fixed $p = 0.75$ and $\tau = 5$ as hyperparameters of the prior distribution \eqref{prior_pos}. We run $R = 400$ replications and the size of the generated chains of the Algorithm \ref{alg:RAM} in each replication was 50,000.  

\begin{figure}[htb!]
    \centering
    \includegraphics[width=0.85\linewidth]{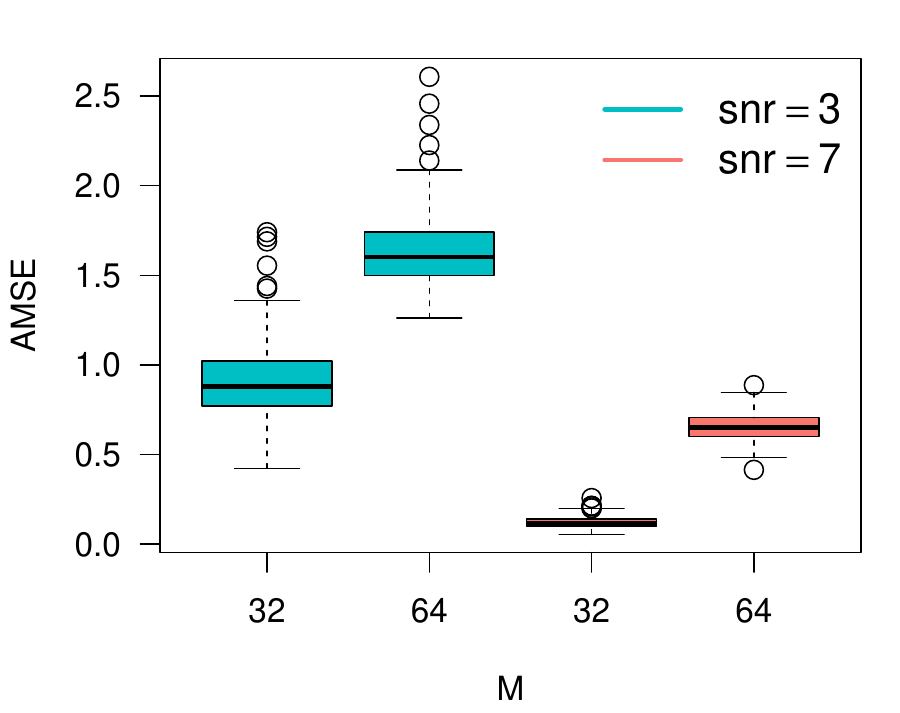}
    \caption{Boxplots of the MSE of the proposed method in the model under iid gamma errors for $L=2$ component functions (Bumps and Doppler). The MSE is related to the estimation of the aggregated curve.}
    \label{fig: bp_sim1_1}
\end{figure}

Table \ref{tab:sim1_alpha} presents the AMSE and the standard deviation of the MSE of the proposed method for each component function, $M$, $L$ and SNR. 
The results indicate a clear deterioration in the method’s performance as the number of component curves $L$ to be estimated increases. Furthermore, the error was consistently smaller in scenarios with higher SNR, as expected, since such settings involve lower noise levels and therefore provide more favorable signal recovery conditions. In addition, Table \ref{tab:sim1_agg} presents the AMSE and the standard deviation of the proposed method but for the aggregated curves $A(\cdot)$ and Figure \ref{fig: bp_sim1_1} shows the boxplots of the MSE for $L=2$ component functions. 

\begin{table}[htb!]
    \centering
    \caption{AMSE (standard deviation) of the proposed method for each component function under gamma noise.}
    \label{tab:sim1_alpha}
    \begin{tabular}{ccccc}\hline\hline
    Function & SNR & M & $L=2$ & $L=4$ \\\hline
    \multirow{4}{*}{Bumps}
        & \multirow{2}{*}{3}
            & 32  & 0.925 (0.279) & 4.235 (1.478) \\ 
            && 64 & 1.637 (0.302) & 4.878 (1.279) \\\cline{2-5}
        & \multirow{2}{*}{7}
            & 32  & 0.121 (0.038) & 0.663 (0.212) \\
            && 64 & 0.652 (0.17) & 1.434 (0.459) \\\hline
            
    \multirow{4}{*}{Doppler}
        & \multirow{2}{*}{3}
            & 32  & 0.877 (0.282) & 4.125 (1.504) \\ 
            && 64 & 1.616 (0.301) & 4.814 (1.333) \\\cline{2-5}
        & \multirow{2}{*}{7}
            & 32  & 0.121 (0.038) & 0.645 (0.245) \\ 
            && 64 & 0.659 (0.163) & 1.389 (0.448) \\\hline

    \multirow{4}{*}{Blocks}
        & \multirow{2}{*}{3}
            & 32  & - & 4.149 (1.463) \\ 
            && 64 & - & 4.801 (1.309) \\\cline{2-5}
        & \multirow{2}{*}{7}
            & 32  & - & 0.639 (0.231) \\ 
            && 64 & - & 1.38 (0.525) \\\hline

    \multirow{4}{*}{Heavisine}
        & \multirow{2}{*}{3}
            & 32  & - & 4.225 (1.617) \\ 
            && 64 & - & 4.847 (1.281) \\\cline{2-5}
        & \multirow{2}{*}{7}
            & 32  & - & 0.64 (0.205) \\ 
            && 64 & - & 1.36 (0.466) \\\hline\hline
    \end{tabular}
\end{table}

\clearpage
\begin{table}[htb!]
    \centering
    \caption{AMSE (standard deviation) of the proposed method under gamma noise.}
    \label{tab:sim1_agg}
    \begin{tabular}{cccc}\hline\hline
        SNR & M & $L=2$ & $L=4$ \\\hline
        \multirow{2}{*}{3} 
            & 32 & 0.901 (0.195) & 4.183 (0.843) \\ 
            & 64 & 1.626 (0.185) & 4.835 (0.689) \\\hline
        \multirow{2}{*}{7} 
            & 32 & 0.121 (0.031) & 0.647 (0.127) \\ 
            & 64 & 0.655 (0.076) & 1.391 (0.173) \\\hline\hline
    \end{tabular}
\end{table}

\subsection{Results for correlated (AR and ARFIMA) errors}
In the simulation of the model under correlated errors, we considered three numbers of equally spaced points in each curve, $M = 512, 1024$ and $2048$, three signal-to-noise ratio values, $\mathrm{SNR} = 3, 5$ and $7$, and a sample size $N = 100$. The error structures analyzed included AR(1) processes, with parameters 
$\phi = 0.25$ and $0.5$ as well as ARFIMA(0, $d$, 0) processes, with parameters 
$d = 0.2$ and $0.4$, allowing the evaluation of both short- and long-range dependence. 

\indent Accordingly, four simulation studies were conducted, considering different combinations of the test functions and $R = 200$ replications for each scenario. In the first simulation, the Blocks and Bumps functions ($L=2$) were used, and the results are presented in Table \ref{sim2_1}. The second simulation includes the Doppler and Heavisine functions ($L=2$ too), with results shown in Table \ref{sim2_2}. The third simulation considers the Blocks, Bumps, and Heavisine functions ($L=3$), with results presented in Table \ref{sim2_3}. Finally, the fourth simulation incorporates all the DJ test functions ($L=4$), with results reported in Table \ref{sim2_4}.

\begin{table}[H]
\centering
\caption{AMSE (standard deviation) of the proposed method in the model under iid normal and correlated errors for $L=2$ component functions (Bumps and Blocks). The AMSE is related to the estimation of the aggregated curve.}
\vspace{0.1cm}
\begin{tabular}{ccccc}
  \hline
\textbf{n} & \textbf{Error structure} & \textbf{SNR = 3} & \textbf{SNR = 5} &  \textbf{SNR = 7} \\ 
  \hline
512 & Normal iid & 1.8915(0.0611) & 1.3931(0.0510) & 1.1544(0.0345) \\ 
  512 & AR(1) - $\phi = 0.25$ & 2.1197(0.0711) & 2.0309(0.0485) & 1.9784(0.0409) \\ 
  512 & AR(1) - $\phi = 0.5$ & 2.2401(0.0804) & 2.1276(0.0705) & 2.0541(0.0590) \\ 
  512 & AR(1) - $\phi = 0.9$ & 2.5710(0.1586) & 2.4646(0.1386) & 2.3776(0.1281) \\ 
  512 & ARFIMA(0,0.2,0) & 2.3302(0.1328) & 2.2050(0.1004) & 2.1276(0.0917) \\ 
  512 & ARFIMA(0,0.4,0) & 2.3624(0.2976) & 2.2339(0.2126) & 2.1899(0.1824) \\ 
  1024 & Normal iid & 1.5072(0.0473) & 1.1258(0.0287) & 0.9419(0.0243) \\ 
  1024 & AR(1) - $\phi = 0.25$ & 2.1078(0.0468) & 2.0381(0.0320) & 1.9912(0.0287) \\ 
  1024 & AR(1) - $\phi = 0.5$ & 2.1957(0.0697) & 2.1103(0.0445) & 2.0624(0.0370) \\ 
  1024 & AR(1) -  $\phi = 0.9$ & 2.5463(0.1216) & 2.4301(0.1112) & 2.3487(0.1008) \\ 
  1024 & ARFIMA(0,0.2,0) & 2.3370(0.1091) & 2.2175(0.0860) & 2.1460(0.0719) \\ 
  1024 & ARFIMA(0,0.4,0) & 2.4114(0.2561) & 2.3379(0.2049) & 2.2768(0.1654) \\ 
  2048 & Normal iid & 1.2015(0.0278) & 0.9158(0.0212) & 0.7475(0.0183) \\ 
  2048 & AR(1) - $\phi = 0.25$ & 2.0777(0.0293) & 2.0321(0.0245) & 1.9988(0.0206) \\ 
  2048 & AR(1) - $\phi = 0.5$ & 2.1394(0.0420) & 2.0830(0.0296) & 2.0495(0.0251) \\ 
  2048 & AR(1) - $\phi = 0.9$ & 2.4511(0.1143) & 2.3397(0.0920) & 2.2742(0.0746) \\ 
  2048 & ARFIMA(0,0.2,0) & 2.2915(0.0866) & 2.2027(0.0729) & 2.1407(0.0603) \\ 
  2048 & ARFIMA(0,0.4,0) & 2.4124(0.2431) & 2.3610(0.1934) & 2.3134(0.1645) \\ 
   \hline
\end{tabular}
\label{sim2_1}
\end{table}

\clearpage
\begin{table}[H]
\centering
\caption{AMSE (standard deviation) of the proposed method in the model under iid normal and correlated errors for $L=2$ component functions (Doppler and Heavisine). The AMSE is related to the estimation of the aggregated curve.}
\vspace{0.1cm}
\begin{tabular}{ccccc}
  \hline
\textbf{n} & \textbf{Error structure} & \textbf{SNR = 3} & \textbf{SNR = 5} &  \textbf{SNR = 7} \\ 
  \hline
512 & Normal iid & 1.1974(0.0755) & 0.7514(0.0412) & 0.5369(0.0332) \\ 
  512 & AR(1) - $\phi = 0.25$ & 4.5979(0.0308) & 4.6023(0.0326) & 4.6117(0.0284) \\ 
  512 & AR(1) - $\phi =  0.5$ & 4.5975(0.0354) & 4.6000(0.0319) & 4.6040(0.0338) \\ 
  512 & AR(1) = $\phi = 0.9$ & 4.6431(0.0562) & 4.6319(0.0499) & 4.6118(0.0445) \\ 
  512 & ARFIMA(0,0.2,0) & 4.5889(0.0462) & 4.5882(0.0430) & 4.5888(0.0368) \\ 
  512 & ARFIMA(0,0.4,0) & 4.5869(0.1581) & 4.5737(0.1249) & 4.5557(0.0985) \\ 
  1024 & Normal iid & 0.8586(0.0467) & 0.5131(0.0301) & 0.3440(0.0244) \\ 
  1024 & AR(1) - $\phi = 0.25$ & 4.6526(0.0301) & 4.6772(0.0274) & 4.6924(0.023) \\ 
  1024 & AR(1) - $\phi = 0.5$ & 4.6390(0.0333) & 4.6529(0.0299) & 4.6648(0.0279) \\ 
  1024 & AR(1) - $\phi = 0.9$ & 4.6463(0.0428) & 4.6372(0.0423) & 4.6359(0.0359) \\ 
  1024 & ARFIMA(0,0.2,0) & 4.6177(0.0436) & 4.6228(0.0358) & 4.6276(0.0316) \\ 
  1024 & ARFIMA(0,0.4,0) & 4.6136(0.1488) & 4.5949(0.1087) & 4.5903(0.0947) \\ 
  2048 & Normal iid & 0.5810(0.0372) & 0.3113(0.0194) & 0.2140(0.0136) \\ 
  2048 & AR(1) - $\phi = 0.25$ & 4.7036(0.0231) & 4.7297(0.0185) & 4.7407(0.0157) \\ 
  2048 & AR(1) - $\phi = 0.5$ & 4.6738(0.0272) & 4.7048(0.0261) & 4.7166(0.0206) \\ 
  2048 & AR(1) -  $\phi = 0.9$ & 4.6492(0.0405) & 4.6468(0.0375) & 4.6509(0.0347) \\ 
  2048 & ARFIMA(0,0.2,0) & 4.6346(0.0409) & 4.6434(0.0347) & 4.6563(0.0335) \\ 
  2048 & ARFIMA(0,0.4,0) & 4.6032(0.1253) & 4.5847(0.1112) & 4.5888(0.0900) \\ 
   \hline
\end{tabular}\label{sim2_2}
\end{table}

\clearpage
\begin{table}[H]
\centering
\caption{AMSE (standard deviation) of the proposed method in the model under iid normal and correlated errors for $L=3$ component functions (Bumps, Blocks and Heavisine). The AMSE is related to the estimation of the aggregated curve.}
\vspace{0.1cm}
\begin{tabular}{ccccc}
  \hline
\textbf{n} & \textbf{Error structure} & \textbf{SNR = 3} & \textbf{SNR = 5} &  \textbf{SNR = 7} \\ 
  \hline
512 & Normal iid & 2.2510(0.1452) & 1.5821(0.0832) & 1.2625(0.0743) \\ 
  512 & AR(1) - $\phi = 0.25$ & 4.4710(0.0957) & 4.3859(0.0706) & 4.3385(0.0670) \\ 
  512 & AR(1) - $\phi = 0.5$ & 4.5650(0.1131) & 4.4768(0.0928) & 4.4242(0.0827) \\ 
  512 & AR(1) - $\phi = 0.9$ & 4.8098(0.1931) & 4.7403(0.1655) & 4.6976(0.1528) \\ 
  512 & ARFIMA(0,0.2,0) & 4.6186(0.1499) & 4.5185(0.1193) & 4.4661(0.1051) \\ 
  512 & ARFIMA(0,0.4,0) & 4.6623(0.4457) & 4.5464(0.2921) & 4.5234(0.2733) \\ 
  1024 & Normal iid & 1.7413(0.0943) & 1.2384(0.0674) & 0.9829(0.0528) \\ 
  1024 & AR(1) - $\phi = 0.25$ & 4.4788(0.0750) & 4.4154(0.0651) & 4.3840(0.0584) \\ 
  1024 & AR(1) - $\phi = 0.5$ & 4.5555(0.0904) & 4.4862(0.0743) & 4.4471(0.0673) \\ 
  1024 & AR(1) - $\phi = 0.9$ & 4.8193(0.1539) & 4.7320(0.1389) & 4.6750(0.1143) \\ 
  1024 & ARFIMA(0,0.2,0) & 4.6416(0.1307) & 4.5708(0.1181) & 4.5146(0.0927) \\ 
  1024 & ARFIMA(0,0.4,0) & 4.6817(0.3863) & 4.6167(0.2983) & 4.5860(0.2627) \\ 
  2048 & Normal iid & 1.3434(0.0686) & 0.9480(0.0472) & 0.7415(0.0399) \\ 
  2048 & AR(1) - $\phi = 0.25$ & 4.4659(0.0595) & 4.4305(0.0578) & 4.4035(0.0499) \\ 
  2048 & AR(1) - $\phi = 0.5$ & 4.5275(0.0726) & 4.4693(0.0615) & 4.4441(0.0564) \\ 
  2048 & AR(1) -  $\phi = 0.9$ & 4.7647(0.1244) & 4.6934(0.1090) & 4.6344(0.1066) \\ 
  2048 & ARFIMA(0,0.2,0) & 4.6310(0.1142) & 4.5605(0.1038) & 4.5163(0.0836) \\ 
  2048 & ARFIMA(0,0.4,0) & 4.6724(0.3380) & 4.6314(0.2855) & 4.6100(0.2228) \\ 
   \hline
\end{tabular}
\label{sim2_3}
\end{table}

\clearpage
\begin{table}[H]
\centering
\caption{AMSE (standard deviation) of the proposed method in the model under iid normal and correlated errors for $L=4$ component functions (all the DJ test functions). The AMSE is related to the estimation of the aggregated curve.}
\vspace{0.1cm}
\begin{tabular}{ccccc}
  \hline
\textbf{n} & \textbf{Error structure} & \textbf{SNR = 3} & \textbf{SNR = 5} &  \textbf{SNR = 7} \\ 
  \hline
512 & Normal iid & 1.9290(0.1516) & 1.3500(0.1011) & 1.0787(0.0759) \\ 
  512 & AR(1) - $\phi = 0.25$ & 3.4031(0.0867) & 3.3381(0.0779) & 3.2957(0.0702) \\ 
  512 & AR(1) - $\phi = 0.5$ & 3.4852(0.1100) & 3.4130(0.0971) & 3.3669(0.0775) \\ 
  512 & AR(1) - $\phi = 0.9$ & 3.6679(0.1802) & 3.6088(0.1558) & 3.5647(0.1421) \\ 
  512 & ARFIMA(0,0.2,0) & 3.5254(0.1504) & 3.4467(0.1249) & 3.4076(0.1127) \\ 
  512 & ARFIMA(0,0.4,0) & 3.6991(0.5873) & 3.5478(0.3846) & 3.4857(0.3324) \\ 
   1024 & Normal iid & 1.4856(0.1084) & 1.0679(0.0744) & 0.8546(0.0590) \\ 
  1024 & AR(1) -  $\phi = 0.25$ & 3.3980(0.0730) & 3.3593(0.0625) & 3.3341(0.0610) \\ 
  1024 & AR(1) - $\phi = 0.5$ & 3.4713(0.0911) & 3.4123(0.0772) & 3.3798(0.0666) \\ 
  1024 & AR(1) - $\phi = 0.9$ & 3.6692(0.1340) & 3.6028(0.1294) & 3.5662(0.1133) \\ 
  1024 & ARFIMA(0,0.2,0) & 3.5424(0.1278) & 3.4800(0.1124) & 3.4401(0.0993) \\ 
  1024 & ARFIMA(0,0.4,0) & 3.6688(0.5341) & 3.5791(0.3675) & 3.5415(0.3052) \\ 
  2048 & Normal iid & 1.1614(0.0735) & 0.8263(0.0533) & 0.6563(0.0420) \\ 
  2048 & AR(1) - $\phi =0.25$ & 3.3893(0.0580) & 3.3692(0.0528) & 3.3474(0.0510) \\ 
  2048 & AR(1) - $\phi = 0.5$ & 3.4388(0.0715) & 3.4049(0.0645) & 3.3797(0.0588) \\ 
  2048 & AR(1) - $\phi = 0.9$ & 3.6304(0.1256) & 3.5619(0.1007) & 3.5335(0.0919) \\ 
  2048 & ARFIMA(0,0.2,0) & 3.5260(0.1066) & 3.4747(0.0993) & 3.4494(0.0859) \\ 
  2048 & ARFIMA(0,0.4,0) & 3.6672(0.4354) & 3.5774(0.3239) & 3.5370(0.2729) \\ 
   \hline
\end{tabular}
\label{sim2_4}
\end{table}

In general, it can be observed that the model is reasonably robust to both short- and long-memory error structures. In general, the mean MSE in the most extreme scenario is between three and four times larger than in the ideal case; however, in absolute numerical terms, this difference remains small.

Moreover, in \cite{johnstone-silerman-1997}, not only was a procedure proposed for estimating the variance, but also a thresholding strategy, namely the universal threshold method. In order to compare this approach with the Bayesian method, an additional simulation study was conducted under the most challenging setting: all four test functions aggregated and errors following an ARFIMA(0, 0.4, 0) process.
Table \ref{sim2_5} presents the comparison between the estimation methods, again based on 200 replications for each scenario. From the table, it can be seen that the differences between the approaches are not substantial; nevertheless, the Bayesian estimator generally yields slightly better results than the universal thresholding method.

\begin{table}[H]
\centering
\caption{AMSE (standard deviation) of the proposed method and the Johnstone and Silverman method in the model under ARFIMA(0, 0.4, 0) errors for $L=4$ component functions (all the DJ test functions). The AMSE is related to the estimation of the aggregated curve.}
\vspace{0.1cm}
\begin{tabular}{cccc}
  \hline
\textbf{n} & \textbf{SNR} & \textbf{Bayesian} & \textbf{Johnstone e Silverman} \\ 
  \hline
512 & 3 & 3.6632(0.5538) & 3.6705(0.5135) \\ 
  512 & 5 & 3.5710(0.4164) & 3.6184(0.3984) \\ 
  512 & 7 & 3.4881(0.3236) & 3.5664(0.3137) \\ 
  1024 & 3 & 3.6906(0.5225) & 3.6952(0.4794) \\ 
  1024 & 5 & 3.5929(0.3683) & 3.6360(0.3545) \\ 
  1024 & 7 & 3.5159(0.2863) & 3.5778(0.2839) \\ 
  2048 & 3 & 3.7140(0.4933) & 3.7239(0.4669) \\ 
  2048 & 5 & 3.5533(0.3115) & 3.5890(0.2983) \\ 
  2048 & 7 & 3.5268(0.2627) & 3.5756(0.2515) \\ 
   \hline
\end{tabular}
\label{sim2_5}
\end{table}

\section{Real data illustrations}

\section{Final considerations}
Within the framework of the Aggregated Functional Data Analysis paradigm, wavelet basis expansions were employed to estimate the mean curves of component functions under positive and correlated error models. To this end, both classical and Bayesian approaches were investigated, along with the necessary computational aspects required for implementation, including the Random Walk Metropolis (RAM) algorithm.

In this sense, it was possible to develop a model for aggregated curves under a positive error structure, specifically, the Gamma distribution—which remains relatively unexplored in the literature and therefore presents clear publication potential. simulation studies were conducted to examine the behavior of the proposed method across a wide range of scenarios, allowing for a comprehensive assessment of its empirical performance.

Additionally, we conducted an extensive simulation study to assess the robustness of the wavelet-based method under violations of the independence assumption for the errors, considering both short and long-memory dependence structures. The results indicate that, although the presence of correlation among the errors increases the mean squared error relative to the ideal independent-error scenario, the estimator’s overall performance remains stable and satisfactory across most of the configurations examined. Furthermore, the comparison between the Bayesian shrinkage rule and the universal thresholding method of Johnstone and Silverman suggests a slight advantage in favor of the Bayesian estimator, particularly under more challenging dependence settings.

\section*{Acknowledgments}
A.R.S.S, V.R.P and R.G.R were financed by Programa de Incentivo a
Novos Docentes (PIND) da Universidade Estadual de Campinas, grant 3376/23 and J.V.S.R was financed by Fundação de Amparo à Pesquisa do Estado de São Paulo (FAPESP), grant 11042/0.


\bibliographystyle{plainnat}
\bibliography{references}

\end{document}